\documentclass[reprint,aps,prb,amsmath,amssymb,floatfix]{revtex4-2}
\usepackage{bm}
\usepackage{graphicx}
\usepackage{booktabs}
\usepackage{braket}
\usepackage{xcolor}
\usepackage[colorlinks,linkcolor=blue,anchorcolor=blue,citecolor=blue,urlcolor=blue]{hyperref}
\begin{document}

\title{Distance Structure of Insertion Kernels and Local Port Response in Non-Reciprocal Su--Schrieffer--Heeger Chains}

\author{Xudong Zhang}
\affiliation{Department of Physics, Wuhan University of Technology, Wuhan 430070, China}

\author{Bin Guo}
\email{binguo@whut.edu.cn}
\affiliation{Department of Physics, Wuhan University of Technology, Wuhan 430070, China}

\date{\today}

\begin{abstract}
Non-reciprocal lattices offer a flexible setting for directional wave transport, raising the question of how enhanced internal scattering translates into transmission between fixed external ports. We address this question for non-reciprocal Su--Schrieffer--Heeger (SSH) chains with staggered gain and loss by relating the reference-dependent insertion kernel $K_p=VG_pV$ to the coherent transmission intensity $T$. On a fixed finite graph, a strong-insertion expansion gives a contribution proportional to $\gamma^{1-d}$, whose coefficient is the signed sum of shortest directed paths of length $d$. When this sum is nonzero, it sets the leading behavior independently of fixed diagonal losses; when it cancels, those losses can enter the first surviving higher-order term. For a fixed cross-sublattice endpoint channel, transmission has a strictly negative slope at every nonzero differentiable stationary point of the kernel magnitude. In the weak-insertion regime, the reference-system coefficient $\chi=-\mathrm{Re}\,\mathrm{Tr}[(G_{0p}Z)^2]$ predicts the initial transmission change for symmetric SSH ports, and $\chi>0$ together with a strictly stable baseline guarantees local stable enhancement. Finite-chain calculations and driven dynamics connect these results to observable response and distinguish asymptotic convergence from transient settling. A strictly dissipative short-chain example exhibits local enhancement without net power gain. Together, these results separate graph-path structure, coherent port response, and dynamical establishment, providing channel-specific criteria for interpreting internal scattering.
\end{abstract}

\maketitle

\section{Introduction}

Non-Hermitian lattices extend wave-transport design by combining gain, loss, and non-reciprocal coupling \cite{Hatano1996,Bender1998,ElGanainy2018,Ashida2020,Bergholtz2021,Miri2019}. Under open boundaries, non-reciprocal hopping can drive the non-Hermitian skin effect (NHSE), concentrating eigenmodes near an edge and modifying conventional bulk-boundary correspondence. Non-Bloch, biorthogonal, and transfer-matrix approaches have clarified this accumulation and its spectral and topological implications \cite{Yao2018,Kunst2018,Yokomizo2019,Okuma2020,Zhang2020,Borgnia2020,KunstDwivedi2019}; biorthogonal entanglement polarization provides a related topological diagnostic \cite{ZhangSunGuo2026}. Photonic and circuit experiments demonstrate controllable gain, loss, and boundary accumulation \cite{Regensburger2012,Weidemann2020,Helbig2020,Hofmann2020}, while quasiperiodic systems offer a further route to non-Hermitian localization \cite{Longhi2019}. These advances motivate a concrete response question: how does the resulting internal structure affect the transmission measured between a specified pair of ports?

Answering this question requires separating spatial accumulation, internal scattering, and coherent port output. A state can be strongly localized near a boundary while the response of a fixed transmission channel remains weak. Conversely, an internal scattering contribution can change the output through coherent interference with the reference channel. For an insertion kernel, the measured amplitude includes both propagation into and out of the kernel and interference with lower-order terms. A further distinction arises in active systems: a regular real-frequency resolvent specifies a formal response, whereas spectral stability determines whether that response is an attracting driven steady state. Added noise, bandwidth, and saturation impose further constraints on device performance \cite{Metelmann2015,Ramos2021,Caves1982,Clerk2010}.

Existing formalisms provide complementary tools for resolving these distinctions. Green-function methods connect bulk spatial growth to boundary response and give explicit finite-chain and multiband expressions \cite{Zirnstein2021,Li2022,Hu2023,Chen2024}. Mechanical, non-Bloch, finite-size, disorder-robust, and singular-value analyses characterize directional and boundary response \cite{Schomerus2020,Xue2021,Liu2023,Wanjura2021,Brunelli2023}. Scattering and input-output theory relate an effective non-Hermitian generator to a measurable channel amplitude \cite{Trivedi2018,Ramos2021}, while weighted-walk expansions describe propagation through finite graphs \cite{Giscard2013} and response derivatives quantify local parameter sensitivity \cite{Lau2018}. Stability analyses then determine whether the driven response is dynamically established \cite{Porras2019,Wanjura2020,Bid2025}. Together, these tools connect internal structure to external response while retaining the conditions needed for each inference.

In this paper we deploy these tools to analyze staggered perturbations on fixed reference systems, addressing two interconnected questions. First, which geometric properties of a finite directed graph determine the kernel's strong-insertion behavior, and how does destructive interference between shortest paths change that behavior? Second, how is the internal insertion kernel related to port-to-port transmission, and can the unperturbed reference system predict the initial output change under weak insertions? We address the path question on finite directed graphs and the port-response question in non-reciprocal SSH chains.

The finite-graph expansion and the local curvature coefficient $\chi$ provide complementary answers, while spectral stability and driven transient dynamics delimit their physical applicability. All comparisons share a fixed frequency baseline and input-output channel definition, with the reference parameters held fixed as $\gamma$ varies. This controlled setting separates three logically distinct layers -- graph-path structure, port-output improvement, and steady-state establishment. We provide an exclusion relation for the cross-sublattice endpoint channel and a reference-system predictor for symmetric SSH ports, and use finite-chain calculations and driven trajectories to examine local stable enhancement and the approach to the steady state.

Section~\ref{sec:model} defines the fixed reference system, the insertion kernel, the observable port output, and the reciprocal representation. Section~\ref{sec:distance} develops the shortest-path expansion and the path-interference mechanism, with an exact four-node example and the SSH end-to-end scaling. Section~\ref{sec:kernel-to-output} connects the internal kernel to the observable port output, proving the exclusion relation and deriving the local predictive coefficient $\chi$. Section~\ref{sec:stability} examines spectral stability, driven convergence, finite-chain validation, and their broader physical implications, including experimental platforms and passivity. Section~\ref{sec:conclusion} concludes.

\section{Model and response decomposition}
\label{sec:model}

We consider an open Su--Schrieffer--Heeger (SSH) dimerized chain \cite{Su1979} of $N$ unit cells, spanned by the local basis $\{\ket{A_0},\ket{B_0},\dots,\ket{A_{N-1}},\ket{B_{N-1}}\}$. Setting $\hbar=1$ and scaling energies by a characteristic coupling $J$, the positive intracell hopping amplitudes are $a=t_1-\delta>0$ and $b=t_1+\delta>0$, with intercell hopping $t_2>0$. The diagonal-free non-Hermitian SSH Hamiltonian and the staggered gain-loss insertion operator are
\begin{subequations}
\label{eq:H0V}
\begin{align}
H_0 &= \sum_{n=0}^{N-1}\!\big(a\ket{A_n}\!\bra{B_n}+b\ket{B_n}\!\bra{A_n}\big) \notag\\
&\quad+ t_2\sum_{n=0}^{N-2}\!\big(\ket{A_{n+1}}\!\bra{B_n}+\ket{B_n}\!\bra{A_{n+1}}\big), \\
V &= i\gamma Z\ (\gamma\ge 0),Z=\sum_{n=0}^{N-1}\big(\ket{A_n}\!\bra{A_n}-\ket{B_n}\!\bra{B_n}\big).
\end{align}
\end{subequations}

Uniform background dissipation $\kappa\ge0$ and boundary port losses are incorporated into the reference operator
\begin{subequations}
\label{eq:Hp}
\begin{align}
H_{0p} &= H_0-i\kappa I-\tfrac{i\Gamma_L}{2}\ket{A_0}\!\bra{A_0}-\tfrac{i\Gamma_R}{2}\ket{B_{N-1}}\!\bra{B_{N-1}}, \\
H_p &= H_{0p}+V.
\end{align}
\end{subequations}

The bare on-site resonance is the zero-energy origin, and $\omega\in\mathbb{R}$ is the physical probe detuning. Holding all hopping rates, decay channels, ports, $\kappa$, and $\omega$ fixed and varying only $\gamma$, the resolvents and the insertion kernel read
\begin{equation}
G_{0p}=(\omega I-H_{0p})^{-1},G_p=(\omega I-H_p)^{-1},K_p=VG_pV.
\label{eq:resolvents}
\end{equation}

The exact scattering $\mathsf T$-operator with respect to $H_{0p}$ satisfies \cite{Economou2006}
\begin{equation}
G_p = G_{0p}+G_{0p}\mathsf T_pG_{0p},\mathsf T_p = V+VG_pV = V+K_p,
\label{eq:Top}
\end{equation}
where $K_p$ is the remainder after subtracting the single-insertion term $V$. This insertion kernel has energy dimensions, depends on the reference system, and is not itself a transmission coefficient. Since $V$ is diagonal, its elements satisfy
\begin{equation}
(K_p)_{ij}=-\gamma^2 Z_{ii}Z_{jj}(G_p)_{ij}.
\label{eq:Kp-elements}
\end{equation}

Cross-sublattice pairs carry a positive sign and intra-sublattice pairs a negative sign. At fixed $\gamma>0$, each kernel element is exactly proportional to the corresponding Green-function element. Their physical roles and extrema as functions of $\gamma$ differ because the proportionality factor contains $\gamma^2$.

Consider a localized port excitation $f=\ket{A_0}$ and detection port $\ell=\ket{B_{N-1}}$. The end-to-end response amplitude $A$, with inverse-energy dimensions, and the dimensionless coherent transmission intensity are
\begin{equation}
A=\ell^\dagger G_p f,T=\Gamma_L\Gamma_R|A|^2.
\label{eq:AT}
\end{equation}
Under broadband, flux-normalized Markovian input-output boundary conditions \cite{Gardiner1985}, defining the port-coupling matrix $W=(\sqrt{\Gamma_L}\ket{A_0},\sqrt{\Gamma_R}\ket{B_{N-1}})$ gives $H_p=H_c-\tfrac{i}{2}WW^\dagger$ with $H_c=H_0+V-i\kappa I$; this makes explicit that the output channel is fixed and that no direct port-to-port channel bypasses the lattice. Equivalently, in the time domain the port variables obey
\begin{equation}
i\dot x = H_p x+Ws_{\rm in},s_{\rm out}=s_{\rm in}-iW^\dagger x,
\label{eq:portvars}
\end{equation}
and the continuous-wave scattering matrix has the Green-function form \cite{FisherLee1981}
\begin{equation}
\mathsf S(\omega)=I_2-iW^\dagger G_pW,\quad t_{RL}=-i\sqrt{\Gamma_R\Gamma_L}\,A,
\label{eq:Sscatt}
\end{equation}
so that $T=|t_{RL}|^2$ \cite{Trivedi2018,Wanjura2020}. If a direct port-to-port channel were present, its coherent amplitude would have to be added explicitly, analogous to the interference between a direct pathway and a resonant channel \cite{Fan2003}. Equation~\eqref{eq:portvars} describes either the classical linear-wave amplitude or the first moment $x=\langle\hat x\rangle$ of a quantum linear system; zero-mean reservoir noise does not alter this mean-value equation, although it still enters the output fluctuations, and the quantum-noise constraints on phase-insensitive amplification cannot be assessed from $T$ alone \cite{Caves1982}. This paper does not include a noise or saturation model: $T$ is not a noisy quantum transmission probability, and the existence of the real-frequency resolvent does not by itself imply asymptotic stability.

Applying the resolvent identity twice yields an exact decomposition that does not require convergence of the Born series,
\begin{equation}
G_p=G_{0p}+G_{0p}VG_{0p}+G_{0p}K_pG_{0p}.
\label{eq:Gp-decomp}
\end{equation}

Projected onto the detection port, $A=A_0+A_1+A_{\ge2}$, with
\begin{subequations}
\label{eq:A-orders}
\begin{align}
A_0 &= \ell^\dagger G_{0p}f,A_1=\ell^\dagger G_{0p}VG_{0p}f, \\
A_{\ge2} &= \ell^\dagger G_{0p}K_pG_{0p}f=\sum_{u,v}(\ell^\dagger G_{0p})_u(K_p)_{uv}(G_{0p}f)_v.
\end{align}
\end{subequations}

The total port transmission accounts for interference across all orders,
\begin{align}
\frac{T}{\Gamma_L\Gamma_R} &= \sum_{\nu\in\{0,1,\ge2\}}|A_\nu|^2 \notag\\
&\quad+ 2\,{\rm Re}\big(A_0^*A_1+A_0^*A_{\ge2}+A_1^*A_{\ge2}\big).
\label{eq:T-total}
\end{align}

Consequently, a single internal kernel element $(K_p)_{ij}$ cannot substitute for the composite amplitude $A_{\ge2}$ or for the full transmission $T$: the complete output further retains interference between the low- and high-order amplitudes. In the perturbatively convergent regime, $A_{\ge2}$ collects contributions from at least two insertion events; outside that regime it remains an exact resummation remainder. Individual amplitude components can be large while the full output is small owing to coherent cancellation.

Following the similarity transformation for non-Hermitian SSH chains \cite{Yao2018}, we define $\bar t=\sqrt{ab}$, $r=\sqrt{b/a}=e^\eta$, and $S\ket{A_n}=r^n\ket{A_n}$, $S\ket{B_n}=r^{n+1}\ket{B_n}$. Replacing the intracell hopping rates by $\bar t$ while preserving the intercell couplings and diagonal on-site terms maps $H_p$ to a reciprocal chain $\bar H_p$. Because $V$ and the port-decay terms are diagonal, they commute with $S$, so
\begin{equation}
H_p=S\bar H_pS^{-1},G_p=S\bar G_pS^{-1},K_p=S\bar K_pS^{-1},
\label{eq:similarity}
\end{equation}
and the same relations apply to the reference system, yielding the directional scaling $A=r^N\bar A$ \cite{Hatano1996,Yao2018}. Relative to the reciprocal state, the right eigenvectors carry an $r^n$ envelope per unit cell -- the exponential weight of the open-chain NHSE. Taking the unit cell as the unit of length, for $r\ne1$ this weight has amplitude length $\xi_{\rm amp}=1/|\ln r|$ and intensity length $\xi_{\rm int}=1/(2|\ln r|)$. This factor can superpose on the localization already present in the reciprocal backbone; by itself it neither proves the existence of topological edge states nor implies $T>1$. Fixing $\bar t$, $t_2$, and the diagonal terms isolates this weight while leaving the spectrum unchanged, whereas fixing $t_1$ and varying $\delta$ instead changes $\bar t$ simultaneously. Related transport theory also separates directional amplification from spectral scattering conditions \cite{GhaemiDizicheh2021}. This representation underlies the real-part spectral bound and the port parity symmetry used below.

\section{Distance structure of the insertion kernel}
\label{sec:distance}

The strong-insertion behavior follows from the correspondence between matrix powers and weighted directed walks \cite{Giscard2013}. Consider a fixed finite graph with $Z=\mathrm{diag}(z_i)$, $z_i=\pm1$, and set $M=\omega I-H_{0p}$. The resolvent admits a Neumann expansion for $\|ZM\|_2/\gamma<1$,
\begin{equation}
G_p=\sum_{n=0}^\infty\Big(-\frac{i}{\gamma}ZM\Big)^n\Big(\frac{i}{\gamma}Z\Big).
\label{eq:Neumann}
\end{equation}

For a fixed finite matrix, this sufficient convergence condition holds for sufficiently large $\gamma$ and establishes the large-$\gamma$ expansion. Let the shortest directed path from $j$ to $i$ ($i\ne j$) have $d_{ij}$ off-diagonal steps. Fewer than $d_{ij}$ factors of $M$ cannot connect these sites, giving
\begin{equation}
(G_p)_{ij}=O(\gamma^{-d_{ij}-1}),(K_p)_{ij}=O(\gamma^{1-d_{ij}}).
\label{eq:scaling-order}
\end{equation}

Let $\mathcal P_d(j\!\to\! i)$ denote the set of shortest paths of length $d=d_{ij}\ge1$, $p=(v_0=j,v_1,\dots,v_d=i)$. Substituting Eq.~\eqref{eq:Neumann} into $K_p=-\gamma^2ZG_pZ$ yields
\begin{subequations}
\label{eq:leading-coeff}
\begin{align}
(K_p)_{ij} &= C_{ij}^{(d)}\gamma^{1-d}+O(\gamma^{-d}), \\
C_{ij}^{(d)} &= i^{d-1}\!\!\sum_{p\in\mathcal P_d(j\to i)}\!\Big[\prod_{r=1}^d(H_{0p})_{v_rv_{r-1}}\Big]\Big[\prod_{r=1}^{d-1}z_{v_r}\Big].
\end{align}
\end{subequations}

Because a shortest path contains no self-loops or repeated vertices, every factor contributing to $C_{ij}^{(d)}$ is an off-diagonal hopping. On-site energies and diagonal losses therefore leave this coefficient unchanged. They enter higher-order corrections, which can become leading when the shortest-path sum vanishes.

To obtain this coefficient, write $K_p=\sum_{n=0}^\infty\gamma^{1-n}B_n$, where $B_n=-i(-i)^nZ(ZM)^n$.
For $n<d$, $(B_n)_{ij}=0$. At $n=d$, every factor crosses an off-diagonal edge, so each contributing walk is a shortest path. The leftmost $Z$ cancels the sign at $i$, leaving only internal-vertex signs. Each off-diagonal factor of $M$ contributes $-(H_{0p})_{v_rv_{r-1}}$, yielding the phase $-i(-i)^d(-1)^d=i^{d-1}$. With $q=\|ZM\|_2/\gamma<1$, the geometric-series tail bounds the remainder as
\begin{equation}
\big|(K_p)_{ij}-C_{ij}^{(d)}\gamma^{1-d}\big|\le\frac{\gamma q^{d+1}}{1-q}=O(\gamma^{-d}).
\label{eq:remainder-bound}
\end{equation}

For a fixed finite matrix with a nonzero shortest-path sum, Eq.~\eqref{eq:leading-coeff} gives the true leading power law. A unique shortest path with nonzero hopping weights guarantees this condition. If different shortest paths cancel, the leading behavior is set by the first nonvanishing coefficient among $B_{d+1},B_{d+2},\ldots$, where fixed diagonal losses can contribute. If all coefficients vanish, including when no directed path connects the sites, the rational-function structure of the finite-matrix inverse makes that element identically zero wherever the inverse exists.

As an exact illustration, consider a four-node graph with $Z={\rm diag}(1,-1,-1,1)$, $\omega=4J$, and forward couplings $(H_{0p})_{10}=(H_{0p})_{20}=(H_{0p})_{31}=J$, $(H_{0p})_{32}=\sigma J$, with internal diagonal energies $\varepsilon_1,\varepsilon_2$. Setting $m_v=\omega-(H_{0p})_{vv}-i\gamma z_v$, direct Gaussian elimination gives
\begin{equation}
(G_p)_{30}=\frac{J^2}{m_3m_0}\Big(\frac{1}{m_1}+\frac{\sigma}{m_2}\Big),C^{(2)}_{30}=-iJ^2(1+\sigma).
\label{eq:fournode}
\end{equation}

For $\sigma=1$, constructive interference yields $\gamma(K_p)_{30}\to-2iJ^2$. For destructive interference with $\sigma=-1$ and identical internal decay $\varepsilon_1=\varepsilon_2$, complete cancellation occurs at all frequencies. If $\varepsilon_1\ne\varepsilon_2$, the shortest-path leading order vanishes and the next-order term emerges,
\begin{equation}
(K_p)_{30}=J^2(\varepsilon_2-\varepsilon_1)\gamma^{-2}+O(\gamma^{-3}).
\label{eq:fournode-next}
\end{equation}

For example, differential losses $\varepsilon_1=-0.3iJ$ and $\varepsilon_2=-0.8iJ$ give $\gamma^2(K_p)_{30}\to-0.5iJ^3$. This example separates the shortest-path coefficient from the first surviving term: fixed diagonal losses leave the former unchanged, but can control the latter once destructive interference removes the shortest-path contribution. The cancellation is analogous in spirit to graphical interference rules for molecular transmission zeros \cite{Markussen2010}, although here it acts on the strong-insertion expansion of the kernel. The example establishes an algebraic asymptotic result; the stability of the corresponding operating regime remains a separate question.

For the end-to-end channel $A_0\!\to\!B_{N-1}$ in the SSH model, the directed path is unique with length $d=2N-1$, and Eq.~\eqref{eq:leading-coeff} gives
\begin{equation}
|(K_p)_{B_{N-1},A_0}| \sim b^Nt_2^{\,N-1}\gamma^{-2(N-1)}.
\label{eq:SSH-scaling}
\end{equation}

\begin{figure}[t]
\centering
\includegraphics[width=\columnwidth]{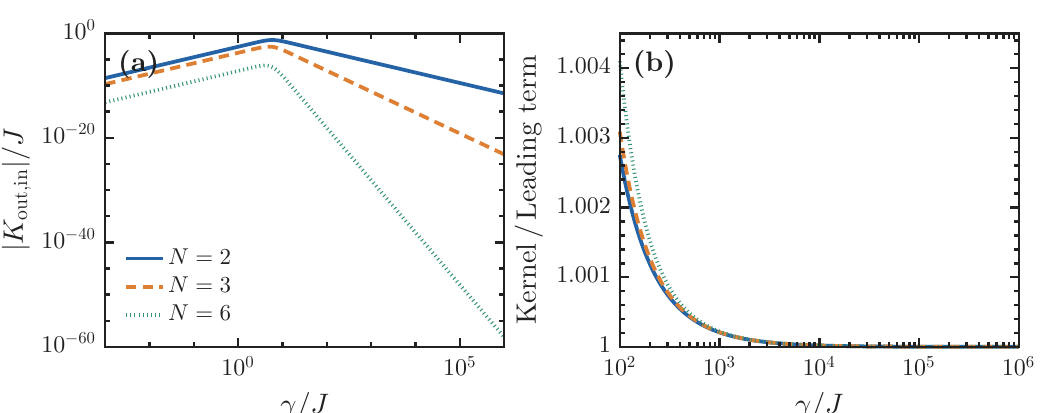}
\caption{Distance scaling of the end-to-end insertion kernel for $t_1/J=1$, $\delta/J=0.4$, $t_2/J=1.6$, $\omega/J=4$, $\kappa/J=4$, $\Gamma_L/J=1$, $\Gamma_R/J=0.6$, and chain lengths $N=2,3,6$. (a) End-to-end kernel magnitude $|(K_p)_{B_{N-1},A_0}|/J$ versus $\gamma/J$, showing a finite-insertion peak followed by the power-law decay of Eq.~\eqref{eq:SSH-scaling}. (b) Ratio of the exact kernel magnitude to the leading-order prediction of Eq.~\eqref{eq:SSH-scaling}, approaching unity for all three chain lengths as $\gamma/J\to\infty$, confirming the asymptotic scaling.}
\label{fig:fig1}
\end{figure}

Figure~\ref{fig:fig1} tests both the distance exponent and the leading coefficient of Eq.~\eqref{eq:SSH-scaling} for $N=2,3,6$. In each case, the kernel grows at small $\gamma$, reaches a finite-insertion maximum, and crosses over to a strong-insertion power-law tail. The unique end-to-end path has length $d=2N-1$, giving tails proportional to $\gamma^{-2}$, $\gamma^{-4}$, and $\gamma^{-10}$ for these three chain lengths. Their increasing steepness thus directly reflects the graph distance between the ports. The ratio of the exact kernel magnitude to Eq.~\eqref{eq:SSH-scaling} approaches unity, testing the leading coefficient as well as the exponent. This growth-to-decay crossover explains the existence of a finite kernel peak; its relation to observable transmission is addressed separately below.

To establish the existence of this finite-insertion peak, note that as $\gamma\to0$, Eq.~\eqref{eq:resolvents} gives $K_p=-\gamma^2ZG_{0p}Z-i\gamma^3ZG_{0p}ZG_{0p}Z+O(\gamma^4)$. Writing the reciprocal Hamiltonian as $\bar H_p=T_{\rm hop}+iD_p$, with $T_{\rm hop}$ real symmetric and $D_p$ real diagonal, the real part of any eigenvalue satisfies
\begin{align}
&{\rm Re}\,E_j=\frac{\psi_j^\dagger T_{\rm hop}\psi_j}{\psi_j^\dagger\psi_j},
\label{eq:realpart-bound1}\\
&\lambda_{\min}(T_{\rm hop})\le{\rm Re}\,E_j\le\lambda_{\max}(T_{\rm hop})\le\bar t+t_2.
\label{eq:realpart-bound2}
\end{align}

Thus $\omega>\bar t+t_2$ guarantees a regular resolvent at the chosen real probe frequency for every $\gamma\ge0$. It does not exclude eigenvalue crossings of the real axis at other frequencies. The tridiagonal endpoint formula gives \cite{Li2022}
\begin{equation}
\begin{aligned}
A&=\frac{C}{\mathcal D_p(\gamma)},\qquad C=b^Nt_2^{\,N-1},\\
\mathcal D_p(\gamma)&=\det(\omega I-H_p).
\end{aligned}
\label{eq:tridiag}
\end{equation}

Because the numerator $C\ne0$ is independent of $\gamma$, for $N\ge2$ the magnitude $|(K_p)_{B_{N-1},A_0}|$ is continuous, vanishes as $\gamma^2$ near $\gamma=0$, and decays as $\gamma^{-2(N-1)}$ as $\gamma\to\infty$; it must therefore attain a global maximum at some finite $\gamma_K>0$. The case $N=1$ corresponds to $d=1$, for which Eq.~\eqref{eq:leading-coeff} gives $C^{(1)}=b$ and $\gamma^{1-d}=1$, so the kernel tends to the constant $b$ -- the constant-tail special case of the distance expansion -- and does not satisfy the same two-sided vanishing condition. This conclusion does not guarantee that the peak is unique, nor does it fix the sign of the eigenvalue imaginary part at the peak.

\section{From internal kernel to observable port output}
\label{sec:kernel-to-output}

Let $F(\gamma)=|(K_p)_{B_{N-1},A_0}|$. From Eq.~\eqref{eq:Kp-elements}, for cross-sublattice boundaries $F(\gamma)=\gamma^2|A(\gamma)|$. Taking logarithmic derivatives of $T=\Gamma_L\Gamma_R|A|^2$ gives
\begin{equation}
\frac{d\ln T}{d\gamma}=2\frac{d\ln F}{d\gamma}-\frac{4}{\gamma}.
\label{eq:logderiv}
\end{equation}

At any non-zero, differentiable stationary point $\gamma_K>0$ of the kernel, $F'(\gamma_K)=0$, so
\begin{equation}
\left.\frac{d\ln T}{d\gamma}\right|_{\gamma_K}=-\frac{4}{\gamma_K}<0.
\label{eq:exclusion}
\end{equation}

Hence, the total transmission $T$ is strictly decreasing at the kernel peak, proving that $\gamma_K$ cannot coincide with an output peak. This exclusion follows directly from the insertion-weight relation; it neither locates the output peak nor requires the output to be monotonic over the entire range. Configurations with an additional direct through-amplitude, multiple source or detection ports, or $\gamma$-dependent port couplings do not satisfy the same derivation and must be reassessed separately.

Equation~\eqref{eq:tridiag} folds the parameter dependence of the end-to-end output into a determinant,
\begin{equation}
T=\frac{\Gamma_L\Gamma_R|C|^2}{|\mathcal D_p|^2}.
\label{eq:T-det}
\end{equation}

Matrix-function differentiation \cite{HighamRelton2014}, using Jacobi's formula and $\partial_\gamma G_p=iG_pZG_p$, gives
\begin{equation}
\frac{\mathcal D_p'}{\mathcal D_p}=-i\,{\rm Tr}(G_pZ),\quad
\Big(\frac{\mathcal D_p'}{\mathcal D_p}\Big)'={\rm Tr}\big[(G_pZ)^2\big].
\label{eq:jacobi}
\end{equation}

Defining $L(\gamma)=\ln T(\gamma)$, the derivatives at $\gamma=0$ evaluate to
\begin{equation}
\begin{aligned}
L'(0)&=-2\,{\rm Im}\,{\rm Tr}(G_{0p}Z),\\
\tfrac12 L''(0)&=-{\rm Re}\,{\rm Tr}\big[(G_{0p}Z)^2\big]\equiv\chi.
\end{aligned}
\label{eq:chi-def}
\end{equation}

For symmetric ports, $\Gamma_L=\Gamma_R>0$, the reciprocal end-to-end reflection operator $P$ obeys
\begin{equation}
P\bar H_p(\gamma)P=\bar H_p(-\gamma),\bar H_p^T=\bar H_p.
\label{eq:parity}
\end{equation}

Spatial reflection exchanges the ports, matrix transposition reverses the propagation direction, and the directional factor $r^N$ is independent of $\gamma$. Thus $A(\gamma)=A(-\gamma)$, which enforces $L'(0)=0$. Exponentiating $L(\gamma)-L(0)=\chi\gamma^2+O(\gamma^4)$ then yields
\begin{equation}
\frac{T(\gamma)}{T(0)}=1+\chi\gamma^2+O(\gamma^4),\chi=-{\rm Re}\,{\rm Tr}\big[(G_{0p}Z)^2\big].
\label{eq:T-chi}
\end{equation}

The coefficient $\chi$ has dimensions of inverse energy squared and is determined entirely by the zero-insertion reference system. For symmetric SSH ports, Eq.~\eqref{eq:T-chi} gives a direct interpretation: $\chi>0$ guarantees enhancement under a sufficiently weak positive insertion, whereas $\chi<0$ guarantees suppression. If $\chi=0$, the next nonvanishing term determines the initial trend. Thus the reference system provides a local predictor of the coherent output without a finite-$\gamma$ search for an optimum.

This interpretation relies on the even endpoint response implied by the reflection and transpose relations in Eq.~\eqref{eq:parity}. Equal port rates alone do not establish that symmetry for an arbitrary network, and the argument does not require the entire kernel matrix to be even in $\gamma$. For asymmetric ports, the first-order coefficient $s_1=L'(0)$ must generally be retained,
\begin{equation}
\frac{T(\gamma)}{T(0)}=1+s_1\gamma+\Big(\chi+\frac{s_1^2}{2}\Big)\gamma^2+O(\gamma^3),
\label{eq:T-asym}
\end{equation}
so a nonzero $s_1$ determines the initial direction of change before the quadratic term contributes.

Stability supplies an independent condition. If the baseline satisfies $\alpha(0)<-\mu$ with $\mu\ge0$, spectral continuity preserves $\alpha(\gamma)<-\mu$ in a sufficiently small neighborhood of $\gamma=0$. For symmetric SSH ports, combining this margin with $\chi>0$ guarantees a local regime of stable enhancement. The criterion determines the initial trend; the width of the enhancement interval, its magnitude, and the global optimum require further analysis. Comparisons with a zero-insertion design whose other parameters have been re-optimized likewise require a separate optimization.

Table~\ref{tab:table1} tests this prediction for a chain of $N=6$ unit cells with $t_1=J$, $\delta=0.4J$, $t_2=1.6J$, $\omega=4J$, $\Gamma_L=\Gamma_R=J$, at four values of the background loss $\kappa$. The second column lists $\chi$ evaluated from the unperturbed reference $G_{0p}$ at $\gamma=0$; the third column reports the fractional transmission change obtained from an independent, direct evaluation at $\gamma=0.01J$. All four parameter sets show an initial trend consistent with the sign of $\chi$ -- transmission decreases for $\kappa/J=0,1$, where $\chi<0$, and increases for $\kappa/J=4,8$, where $\chi>0$ -- and every small-step probe lies within the asymptotically stable region established in Sec.~\ref{sec:stability}. The reported changes are small by construction, since they test the local coefficient rather than demonstrate a significant device gain.

\begin{table}[t]
\caption{Reference-system prediction versus direct small-step transmission for $N=6$, $t_1=J$, $\delta=0.4J$, $t_2=1.6J$, $\omega=4J$, $\Gamma_L=\Gamma_R=J$. $\chi$ is computed from the unperturbed reference $G_{0p}$ at $\gamma=0$; the third column is obtained from an independent direct evaluation at $\gamma=0.01J$.}
\label{tab:table1}
\begin{ruledtabular}
\begin{tabular}{cccc}
$\kappa/J$ & $\chi J^2$ & $[T(0.01J)/T(0)-1]\times10^6$ & Trend \\
\hline
0 & $-0.939944$ & $-93.989562$ & Decreasing \\
1 & $-0.662923$ & $-66.289988$ & Decreasing \\
4 & $\phantom{-}0.040129$ & $\phantom{-}4.012901$ & Increasing \\
8 & $\phantom{-}0.090699$ & $\phantom{-}9.069928$ & Increasing \\
\end{tabular}
\end{ruledtabular}
\end{table}

For an isolated reciprocal chain $\bar H_c=T_{\rm hop}+i\gamma Z-i\kappa I$, chiral symmetry $\{T_{\rm hop},Z\}=0$ implies that the eigenvalues of $T_{\rm hop}$ occur in pairs $\pm\epsilon_j$ ($\epsilon_j>0$). The determinant and its curvature coefficient evaluate to
\begin{subequations}
\label{eq:chiral}
\begin{align}
\mathcal D_c(\gamma) &= \prod_{j=1}^N\Big[(\omega+i\kappa)^2+\gamma^2-\epsilon_j^2\Big], \\
\chi_c &= -2\sum_{j=1}^N\frac{\omega^2-\kappa^2-\epsilon_j^2}{(\omega^2-\kappa^2-\epsilon_j^2)^2+4\omega^2\kappa^2}.
\end{align}
\end{subequations}

For $\omega>\epsilon_{\max}$, $\kappa^2<\omega^2-\epsilon_{\max}^2$ guarantees $\chi_c<0$, while $\kappa^2>\omega^2-\epsilon_{\min}^2$ guarantees $\chi_c>0$. Background loss can therefore reverse the curvature by changing the real part of the resolvent determinant. This control example gives a mechanism for the loss-dependent sign change also observed in Table~\ref{tab:table1}.

Here $\chi_c$ describes the relative change in the squared modulus of the port-free end-to-end Green's function. A measurable transmission with finite port coupling requires the coefficient in Eq.~\eqref{eq:chi-def}; at zero port coupling, the transmission intensity itself vanishes. The bounds for this isolated-chain example therefore serve as an interpretation of the mechanism and do not directly establish bounds for finite ports.

\begin{figure*}[tp]
\centering
\includegraphics[width=\textwidth]{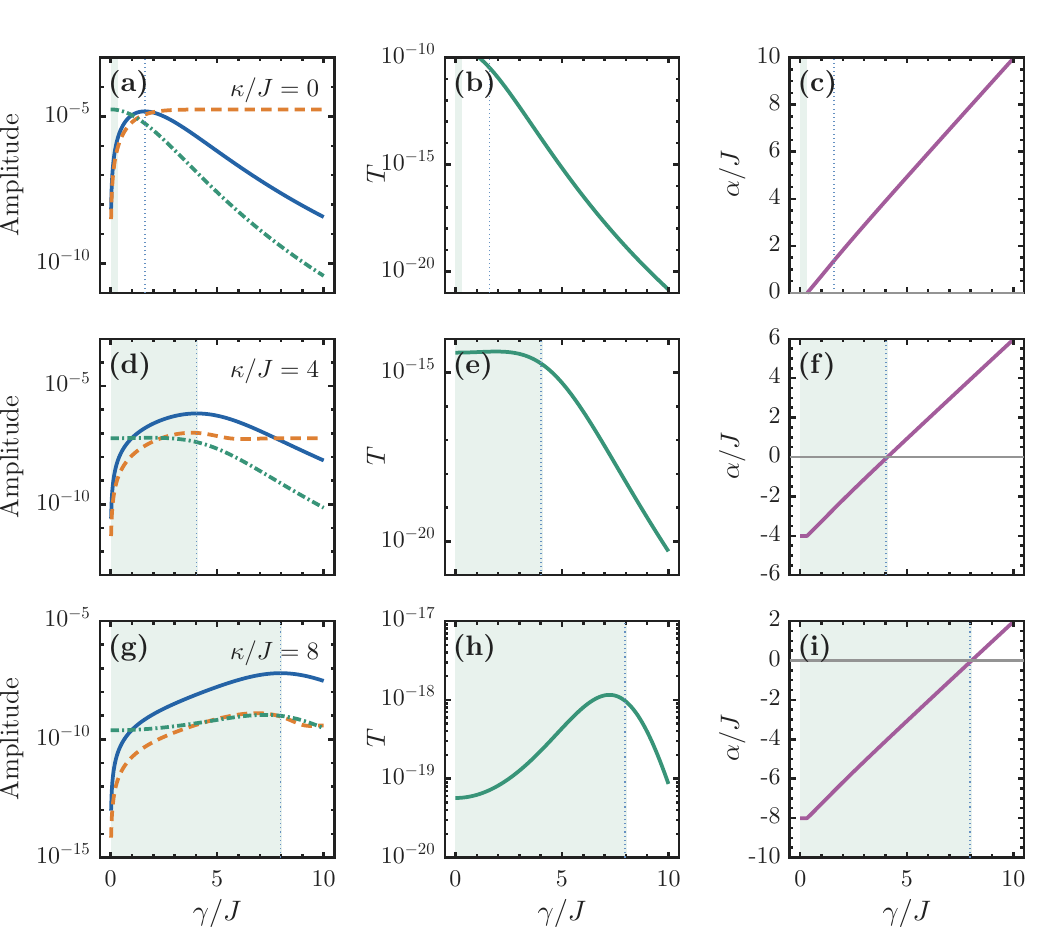}
\caption{Insertion kernel, transmission, and spectral stability for the parameters in Table~\ref{tab:table1}. Panels (a--c), (d--f), and (g--i) use $\kappa/J=0,4,8$, respectively. (a,d,g) Blue solid, orange dashed, and green dash-dotted curves show $|(K_p)_{B_5,A_0}|/J$, $J|A_{\ge2}|$, and $J|A|$. (b,e,h) Transmission intensity $T$. (c,f,i) Spectral stability measure $\alpha/J$. Green shading indicates $\alpha<0$; dotted vertical lines mark $\gamma_K$. The transmission maxima in (e,h) precede the corresponding kernel peaks. Responses outside the stable domain are formal frequency-domain values, not attracting steady states. Exact zero amplitudes at $\gamma=0$ are omitted on logarithmic axes.}
\label{fig:fig2}
\end{figure*}

\section{Spectral stability, finite-chain validation, and physical implications}
\label{sec:stability}

\subsection{Stability, driven dynamics, and numerical profiles}

Asymptotic stability requires the maximum imaginary part of the spectrum to remain negative,
\begin{equation}
\alpha(\gamma,\kappa)=\max_j\,{\rm Im}\,E_j(H_p)<0.
\label{eq:stability}
\end{equation}

Under homogeneous loss, $\alpha(\gamma,\kappa)=\alpha(\gamma,0)-\kappa$ and $G_p(\omega;\kappa)=G_p(\omega+i\kappa;0)$. Because loss both shifts the stability boundary and alters the fixed real-frequency response, the kernel peak for each loss curve must be determined independently.

Under continuous-wave driving $i\partial_t\psi=H_p\psi+fe^{-i\omega t}$ with $\psi(0)=0$, define $\phi(t)=e^{i\omega t}\psi(t)$ and the formal steady state $\phi_{\rm ss}=G_pf$. Their difference obeys
\begin{equation}
\phi(t)-\phi_{\rm ss}=-e^{-i(H_p-\omega I)t}\phi_{\rm ss}.
\label{eq:evolution}
\end{equation}
We quantify state-space and port relative deviations as
\begin{equation}
e_{\rm state}(t)=\frac{\|\phi(t)-\phi_{\rm ss}\|_2}{\|\phi_{\rm ss}\|_2},
e_{\rm out}(t)=\frac{|\ell^\dagger[\phi(t)-\phi_{\rm ss}]|}{|\ell^\dagger\phi_{\rm ss}|}.
\label{eq:errors}
\end{equation}

These two error measures use different denominators, so an extremely small steady-state port amplitude can amplify the relative port error. Spectral stability and transient amplification differ in non-normal systems \cite{Trefethen1993,Trefethen2005,Makris2014}; skin effects can also enhance nonnormality \cite{Nakai2024}. This normalized error is not a statement about total energy gain: the quantity $1/(-\alpha)$ only sets the asymptotic exponential time scale, and the settling time to a given accuracy further depends on the modal weights and the transient dynamics.

For Table~\ref{tab:table1}, $\bar t+t_2\simeq2.5165J<\omega$ excludes probe-frequency poles, so the smooth kernel peaks are not pole divergences. Responses are obtained from finite linear solves, stability from eigenvalues, and dynamics from Eq.~\eqref{eq:evolution}. Extrema are bracketed on $0\le\gamma/J\le10$ and refined locally; stability thresholds solve $\alpha=0$.

Figure~\ref{fig:fig2} and Table~\ref{tab:table2} compare the kernel response, coherent transmission, and spectral stability within the same fixed channel. For $\kappa/J=0,1$, the kernel peaks lie in the unstable region; for $\kappa/J=4,8$, they lie in the stable region. In the latter two cases, transmission reaches its maximum within the scanned interval at $\gamma/J\simeq1.905589$ and $7.228670$, respectively, before the corresponding kernel peaks. This numerical ordering is consistent with Eq.~\eqref{eq:exclusion}, whose analytic content is the negative transmission slope at a nonzero kernel stationary point. The profiles illustrate the distinction between a kernel maximum and a transmission optimum; the general ordering of extrema requires information beyond the exclusion relation.

\begin{table}[t]
\caption{Kernel peaks, output at the peaks, and stability thresholds for the parameters in Table~\ref{tab:table1}. Peaks are searched over $0\le\gamma/J\le10$; $\gamma_s$ solves $\alpha(\gamma_s)=0$. The sign of $\alpha(\gamma_K)$ determines stability.}
\label{tab:table2}
\begin{ruledtabular}
\begin{tabular}{ccccc}
$\kappa/J$ & $\gamma_K/J$ & $\gamma_s/J$ & $\alpha(\gamma_K)/J$ & $T(\gamma_K)$ \\
\hline
0 & 1.600840 & 0.329076 & $\phantom{-}1.385115$ & $3.3144\times10^{-11}$ \\
1 & 1.825164 & 1.248365 & $\phantom{-}0.627964$ & $8.1696\times10^{-12}$ \\
4 & 4.033089 & 4.098801 & $-0.067208$ & $1.8178\times10^{-15}$ \\
8 & 7.974864 & 8.051886 & $-0.077508$ & $9.6470\times10^{-19}$ \\
\end{tabular}
\end{ruledtabular}
\end{table}

The small outputs in Table~\ref{tab:table2} reflect the chosen chain length, loss, and off-band detuning. A short-chain example shows that stable enhancement can also occur with appreciable coherent transmission. Consider $N=2$, $\kappa=0.5J$, $\omega=0$, with the hopping and port rates of Table~\ref{tab:table1}. For $\tilde\gamma=\gamma/J$, the determinant is $\mathcal D_p/J^4=4.3556-5.49\tilde\gamma^2+\tilde\gamma^4$, giving $\chi J^2=2.520893$. At $\gamma=0.2J$, $T$ rises from $0.518389$ to $0.574454$, approximately $10.8\%$, while $\alpha/J=-0.583750$. The interval $0\le\gamma/J\le0.2$ satisfies the strict passivity condition below. This example establishes local stable enhancement within a strictly dissipative regime.

\begin{figure}[tbp]
\centering
\includegraphics[width=\columnwidth]{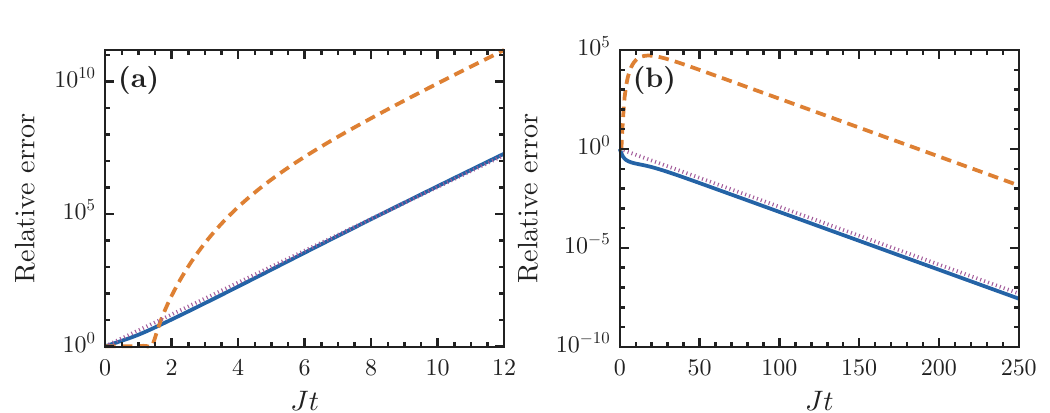}
\caption{Driven convergence at kernel peaks, with parameters as in Table~\ref{tab:table1}. Blue solid and orange dashed curves show $e_{\rm state}$ and $e_{\rm out}$ from Eq.~\eqref{eq:errors}; purple dotted curves show $e^{\alpha t}$ as rate guides. (a) Unstable peak at $\kappa=0$, $\gamma=1.6008J$. The errors eventually grow, but need not increase monotonically. (b) Stable peak at $\kappa=4J$, $\gamma=4.0331J$. Both errors converge to zero after a transient. The distinct normalizations preclude interpreting their relative heights as energy gain.}
\label{fig:fig3}
\end{figure}

Figure~\ref{fig:fig3} compares driven trajectories at an unstable and a stable kernel peak. In the unstable example, both error measures eventually grow, so the formal frequency-domain response is not an attracting steady state. At the stable peak, the state and port errors ultimately decay to zero after a pronounced transient in the port response. The curves $e^{\alpha t}$ indicate exponential growth or decay rates and are not bounds on either error. A large relative port deviation can arise from modal interference and a small steady-state denominator, so these trajectories alone do not isolate non-normal energy amplification. Spectral stability determines eventual convergence, whereas modal weights and transient dynamics determine the approach to the steady state and the settling time to a prescribed accuracy.

\subsection{Relation to path-sum and response-derivative methods}

The two analytic results connect established methods to the insertion problem in a fixed channel. Equation~\eqref{eq:leading-coeff} uses weighted-walk methods \cite{Giscard2013} to resolve the staggered kernel into path contributions. Equation~\eqref{eq:T-chi} combines the endpoint determinant with reference-response derivatives \cite{Li2022,Lau2018} to predict the initial output trend. Related logarithmic Green-function derivatives probe spectral winding under boundary variation \cite{LiMu2021}. The present path expansion applies to fixed finite graphs, including loops, whereas the SSH similarity transformation and even-output criterion rely on the chain geometry, fixed endpoint numerator, and reflection/transpose relations. Applying the output criterion to other networks therefore requires checking those conditions explicitly.

These results complement non-Hermitian response theory \cite{Bid2025} and directional-amplification frameworks \cite{Porras2019,Wanjura2020,Metelmann2015,McDonald2018}. Graph-based \cite{Ranzani2015} and coupled-mode \cite{Fan2003} descriptions likewise connect internal dynamics to specified external channels. Non-normal dynamics \cite{Trefethen2005}, quantum amplifier noise \cite{Caves1982,Clerk2010}, and quantum sensing \cite{McDonaldClerk2020} introduce performance questions beyond the initial coherent transmission trend. An off-exceptional-deficiency sensing study illustrates how noise can change the preferred operating regime \cite{LiGuo2026}. Assessing bandwidth, added noise, saturation, or sensing advantage requires the corresponding dynamical and noise models \cite{Metelmann2015,Aspelmeyer2014,Caves1982,Clerk2010,Malz2018}.

\subsection{Experimental platforms and passivity}
\label{sec:platforms}

Synthetic photonic lattices realize controllable gain and loss \cite{Regensburger2012,Weidemann2020}, while active topolectrical circuits realize non-reciprocal coupling \cite{Helbig2020}. Reciprocal skin effects have also been demonstrated in passive circuits \cite{Hofmann2020}, through a mechanism distinct from the present non-reciprocal chain. Feedback-controlled mechanical metamaterials \cite{Brandenbourger2019,Ghatak2020} and acoustic lattices \cite{ZhangAcoustic2021} realize non-reciprocal interactions. Pumped Josephson circuits demonstrate directional amplification \cite{Sliwa2015}. Optomechanical experiments realize reconfigurable non-reciprocity and isolation \cite{Shen2018,Fang2017,Bernier2017}; corresponding theories address coupling mechanisms and added noise \cite{Aspelmeyer2014,Malz2018}. These works establish individual model ingredients, not a realization of our full parameter set. Port normalization, effective bandwidth, and the mapping from experimental response matrices require platform-specific calibration.

Net gain in an active implementation requires pumping or external feedback. The condition $\gamma\le\kappa$ excludes net on-site gain but does not ensure passivity of the full coupling matrix. With energy normalized as $\|x\|_2^2$, the bare dissipation matrix $D=i(H_c-H_c^\dagger)$ has eigenvalues $2\kappa\pm2\sqrt{\gamma^2+\delta^2}$ per unit cell. Thus $\kappa\ge\sqrt{\gamma^2+\delta^2}$ ensures $D\succeq0$ and non-increasing undriven energy; the ports add $WW^\dagger\succeq0$. The strict condition $\kappa>\sqrt{\gamma^2+\delta^2}$ directly guarantees energy decay and asymptotic stability. At equality, any undamped subspace must be checked separately. Such generators can be interpreted using coherent and dissipative couplings \cite{Metelmann2015}, not necessarily reciprocal passive components alone. The short-chain example satisfies the strict condition, so local transmission enhancement does not require net power amplification.

We emphasize three limitations. First, the asymptotic expansion is derived for a fixed finite graph and is not uniform in chain length or under simultaneous parameter variation. Second, the local criterion $\chi$ is a weak-insertion statement: it does not locate global optima or guarantee monotonicity. Third, the transmission intensity $T$ is a coherent mean-value quantity. These limitations define clear directions for future work.

\section{Conclusion}
\label{sec:conclusion}

We have related the internal insertion kernel to directed-path structure and observable port transmission in non-reciprocal SSH chains. In the strong-insertion regime of a fixed finite graph, the signed sum of shortest directed paths determines the leading coefficient. Fixed diagonal losses leave this coefficient unchanged; when path interference cancels it, they can enter the first surviving higher-order contribution. The endpoint insertion-weight relation then gives an exact distinction between internal and external response: transmission decreases at every nonzero differentiable stationary point of the kernel magnitude, excluding that point as an output extremum.

In the weak-insertion regime, the reference system provides a complementary predictor. For symmetric SSH ports, $\chi=-\mathrm{Re}\,\mathrm{Tr}[(G_{0p}Z)^2]$ determines the initial transmission trend, and $\chi>0$ together with a strictly stable baseline guarantees a local interval of stable enhancement. Finite-chain calculations and driven dynamics show how this response is established: spectral stability guarantees asymptotic convergence, while modal weights and transient dynamics shape the approach and the settling time. The strictly dissipative short-chain example further demonstrates that local transmission enhancement can occur without net power gain.

These results organize the response problem into three distinct layers: the graph-path structure of an internal scattering quantity, the coherent output of a specified external channel, and the dynamical establishment of that output. This separation provides operational criteria for interpreting internal scattering and predicting local response.

\section*{Acknowledgments}
This work is supported by the National Natural Science Foundation of China (Grant No. 11975175). 

The authors have no conflicts to disclose.

\section*{Data Availability}
There are no publicly available research data or software supporting this manuscript. Requests for further information or data should be sent to the authors.

\bibliography{references_main}

\end{document}